\documentclass[aps,apl,twocolumn,superscriptaddress]{revtex4}

\usepackage{graphicx}
\usepackage{epstopdf}
\usepackage{comment}
\usepackage{amsmath,amssymb}
\usepackage{bm}
\newcommand{\ctext}[1]{\raise0.2ex\hbox{\textcircled{\scriptsize{#1}}}}

\begin{document}

\title{Phononic-crystal cavity magnomechanics}

\author{D. Hatanaka}

%\email{daiki.hatanaka@ntt.com}
\email{daiki.hatanaka.hz@hco.ntt.co.jp}

\author{M. Asano}

\author{H. Okamoto}

\author{H. Yamaguchi}

\affiliation{NTT Basic Research Laboratories, NTT Corporation, Atsugi-shi, Kanagawa 243-0198, Japan}

\begin{abstract}
Establishing a way to control magnetic dynamics and elementary excitations (magnons) is crucial to fundamental physics and the search for novel phenomena and functions in magnetic solid-state systems. Electromagnetic waves have been developed as means of driving and sensing in magnonic and spintronics devices used in magnetic spectroscopy, non-volatile memory, and information processors \cite{jungwirth2012spin, chumak2015magnon, tabuchi2015coherent, li2020hybrid, awschalom2021quantum}. However, their millimeter-scale wavelengths and undesired cross-talk have limited operation efficiency and made individual control of densely integrated magnetic systems difficult. Here, we utilize acoustic waves (phonons) to control magnetic dynamics in a miniaturized phononic crystal micro-cavity and waveguide architecture. We demonstrate acoustic pumping of localized ferromagnetic magnons, where their back-action allows dynamic and mode-dependent modulation of phononic cavity resonances. The phononic crystal platform enables spatial driving, control and read-out of tiny magnetic states and provides a means of tuning acoustic vibrations with magnons. This alternative technology enhances the usefulness of magnons and phonons for advanced sensing, communications and computation architectures that perform transduction, processing, and storage of classical and quantum information \cite{chumak2015magnon, awschalom2021quantum}.
\end{abstract}  
\maketitle
\begin{figure*}[t]
	\begin{center}
		\vspace{-0.5cm}\hspace{-0.0cm}
		\includegraphics[scale=1]{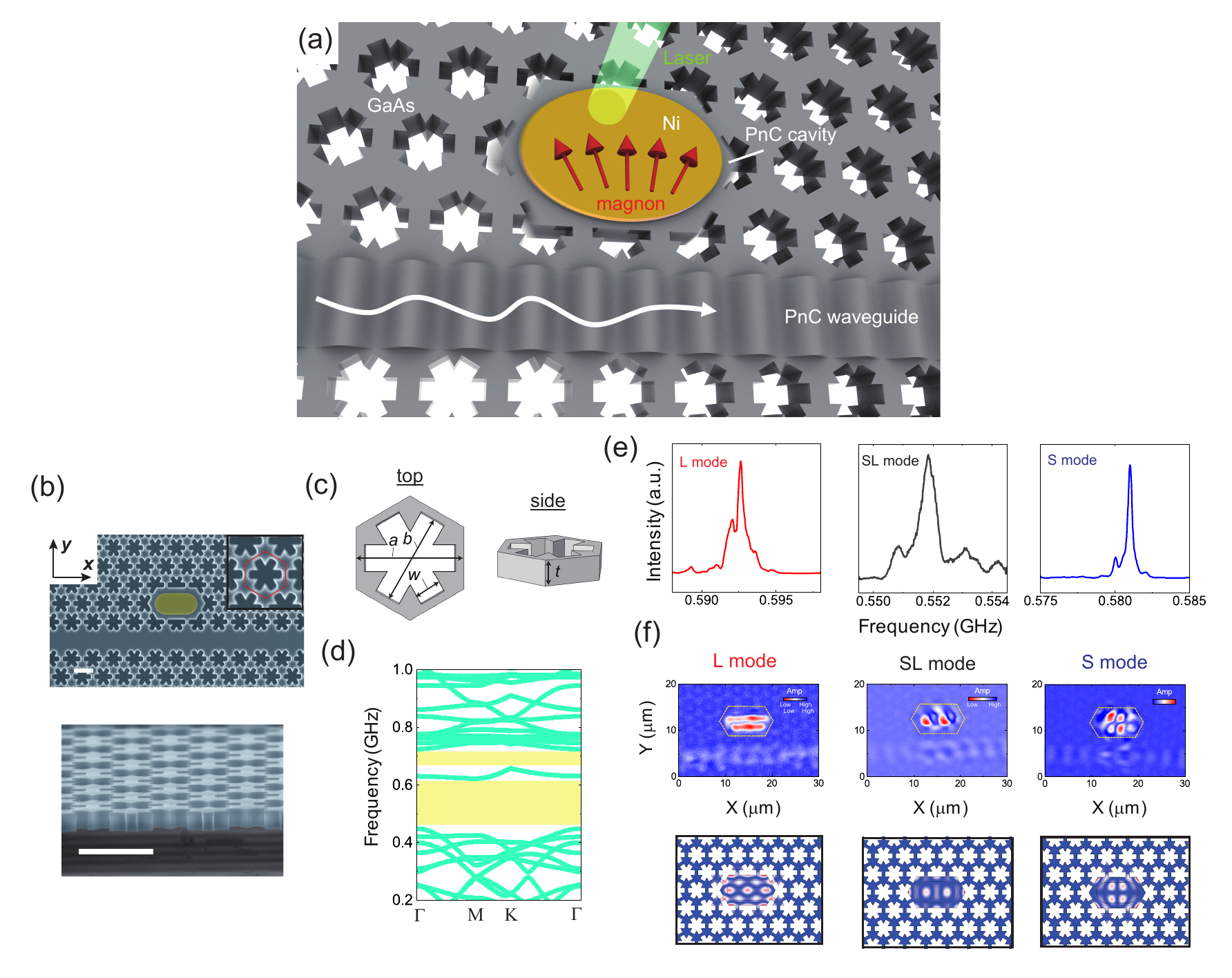}
		\vspace{0.cm}
		\caption{PnC cavity magnomechanics. \textbf{(a)} Schematic diagram of PnC magnomechanical system and measurement configuration. The device consists of a GaAs/Al$_{0.7}$Ga$_{0.3}$As heterostructure and the PnC is formed in a free-standing GaAs membrane. Rayleigh SAWs are excited by applying an alternating voltage via the piezoelectric effect to inter-digit transducers (IDT) and they transform into asymmetric Lamb waves in the PnC membrane. The acoustic waves excite a PnC cavity resonance through a line-defect waveguide, and the resonant vibrations are measured by an optical interferometer at room temperature and in a moderate vacuum. A ferromagnetic Ni film deposited on the cavity sustains magnon oscillations that are driven by cavity vibrations via magnetostriction. \textbf{(b)} Electron microscope images of the magnomechanical cavity-waveguide geometry (top) and cross-section (bottom). The scale bar is 4 $\mu$m. The cavity is formed by removing two air holes from the phononic lattice. The Ni film on the defect is false-colored in yellow. The inset zooms in on the snowflake lattice, and the red dashed line denotes the unit structure. \textbf{(c)} Unit cell of PnC lattice designed with $a=$ 4.0 $\mu$m, $b=$ 3.6 $\mu$m, $w=$ 1.0 $\mu$m and $t=$ 1.0 $\mu$m. \textbf{(d)} Dispersion relation calculated by FEM (COMSOL Multiphysics). A complete phononic bandgap forms between 0.5 and 0.8 GHz, except around 0.61-0.62 GHz (highlighted in yellow). \textbf{(e)} Spectral response of PnC cavity measured with an optical interferometer. From left to right: well-defined acoustic resonance peaks appear at 0.593 GHz (L mode), 0.552 GHz (SL mode), and 0.581 GHz (S mode) with quality-factors $Q_{a} =$ 550, 680 and 1,540. \textbf{(f)} Experimental (top) and simulated (bottom) modal shape of displacement amplitude of L, SL and S modes in the left, middle and right panels.}
		\label{fig 1}
		\vspace{-0cm}
	\end{center}
\end{figure*}
\begin{figure*}[t]
	\begin{center}
		\vspace{-0.5cm}\hspace{-0.0cm}
		\includegraphics[scale=1]{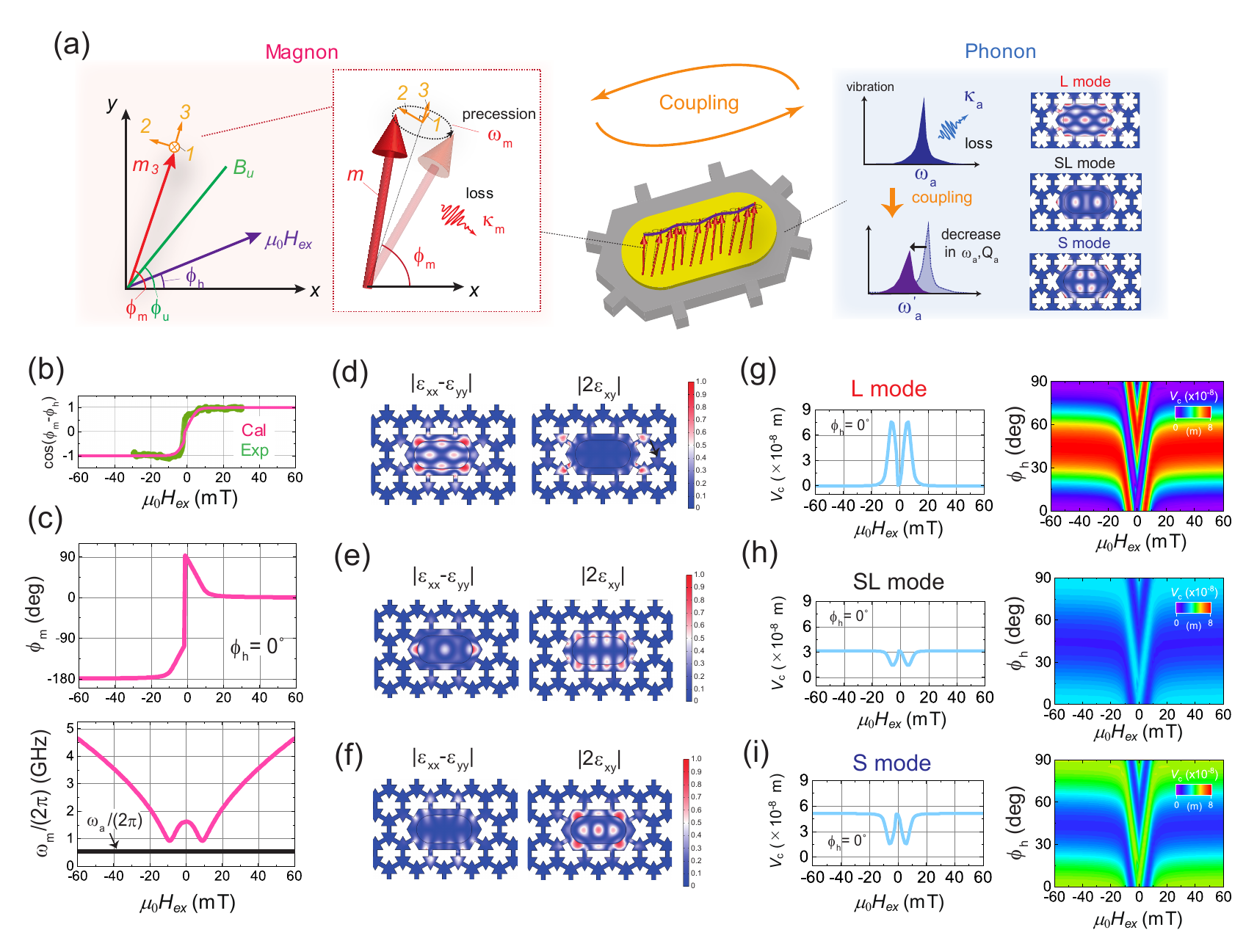}
		\vspace{0.5cm}
		\caption{Magnon-phonon interaction. \textbf{(a)} Left: $xy$-coordinate system in which the external magnetic field ($\mu_{0} H_{\rm ex}$), magnetization ($m_{3}$), and in-plane anisotropic filed ($B_{\rm u}$) respectively have an angle of $\phi_{\rm h}$, $\phi_{\rm m}$ and $\phi_{\rm u}$ away from the $x$-axis. The inset shows the alternative $1$-, $2$- and $3$-axis coordinate system. The $1$- and $2$-axis define the magnetization precession plane normal to the $3$-axis. The spin oscillation with frequency $\omega_{\rm m}$ and damping $\kappa_{\rm m}$ is sustained by the Ni thin film. Right: Acoustic resonant vibrations in the PnC cavity whose frequency and quality factor change due to magnetostrictive coupling. \textbf{(b)} Equilibrium magnetization component projected to the field orientation at $\phi_{\rm h}$ = 0$^{\circ}$ plotted as a function of the bias field ($\mu_{0} H_{\rm ex}$); the green and pink solid lines are experimental and calculated results. The magnetization versus $\mu_{0} H_{\rm ex}$ curve was obtained by magneto-optical Kerr microscopy. \textbf{(c)} (top): Field dependence of the magnetization angle ($\phi_{\rm m}$), (bottom): spin-wave resonance frequency ($\omega_{\rm m} / (2\pi)$). The black solid line in the bottom panel is the acoustic resonant frequency ($\omega_{\rm a} / (2\pi)$). \textbf{(d)-(f)} Spatial distribution of strain components $\left| \epsilon_{\rm xx} - \epsilon_{\rm yy} \right|$ (left) and $\left| 2\epsilon_{\rm xy} \right|$ (right) in L (0.593 GHz, (d)), SL (0.552 GHz, (e)), and S (0.581 GHz) modes simulated by FEM. The color scales of the SL and S modes are normalized by the maximum strain of $\left| 2\epsilon_{\rm xy} \right|$, and that of the L mode by $\left| \epsilon_{\rm xx} - \epsilon_{\rm yy} \right|$. \textbf{(g)-(i)} Left: Field response of magnetoelastic mode coupling ($V_{\rm c}$) at $\phi_{\rm h}$ = 0$^{\circ}$. Right: $\phi_{\rm h}$ dependence of $V_{\rm c}$ versus $\mu_{0} H_{\rm ex}$ response.}
		\label{fig 2}
		\vspace{-0cm}
	\end{center}
\end{figure*}
\begin{figure}[t]
	\begin{center}
		\vspace{-0.2cm}\hspace{-0.0cm}
		\includegraphics[scale=1]{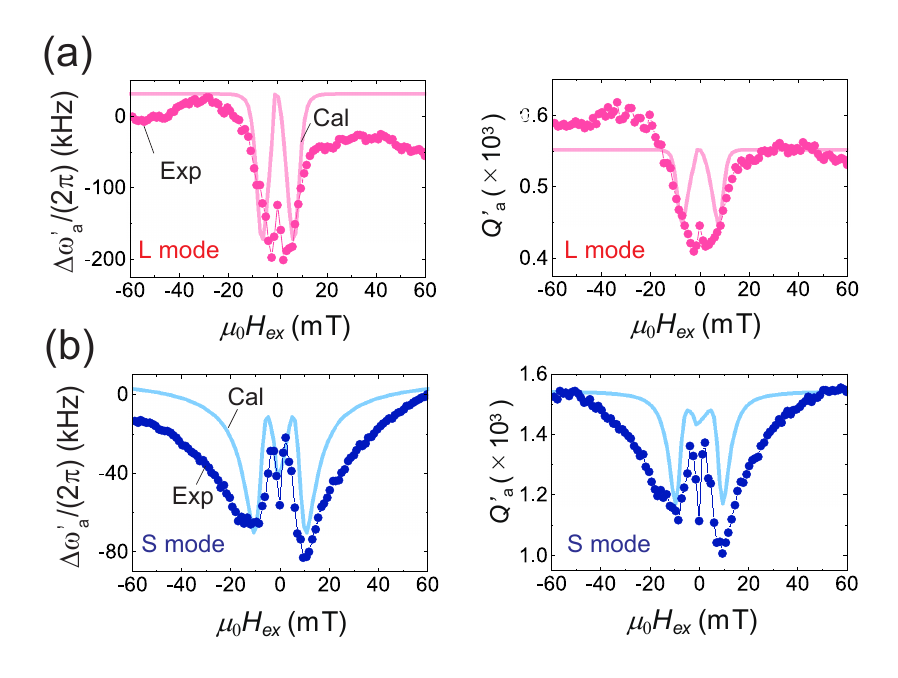}
		\vspace{0.0cm}
		\caption{Magnetoelastic modulation of acoustic cavity resonances. \textbf{(a)} and \textbf{(b)} Field ($\mu_{0} H_{\rm ex}$) response of acoustic resonance frequency shift $\Delta \omega^\prime_{\rm a}/ (2\pi)$ (left) and quality factor $Q_{a}^{\prime}$ (right) in L and S modes. Here, $\Delta \omega^\prime_{\rm a} = \omega^\prime_{\rm a} - \omega_{\rm a0}$ where $\omega_{\rm a0} / (2\pi)$ is the frequency at $\left|\mu_0 H_{\rm ex} \right|$ = 60 mT. The solid circles are experimental data taken from a Lorentzian fitting to the measured cavity spectra. The solid lines are theoretical results based on equation (3).}
		\label{fig 3}
		\vspace{-0.5cm}
	\end{center}
\end{figure}
\begin{figure*}[t]
	\begin{center}
		\vspace{-0.2cm}\hspace{-0.0cm}
		\includegraphics[scale=1]{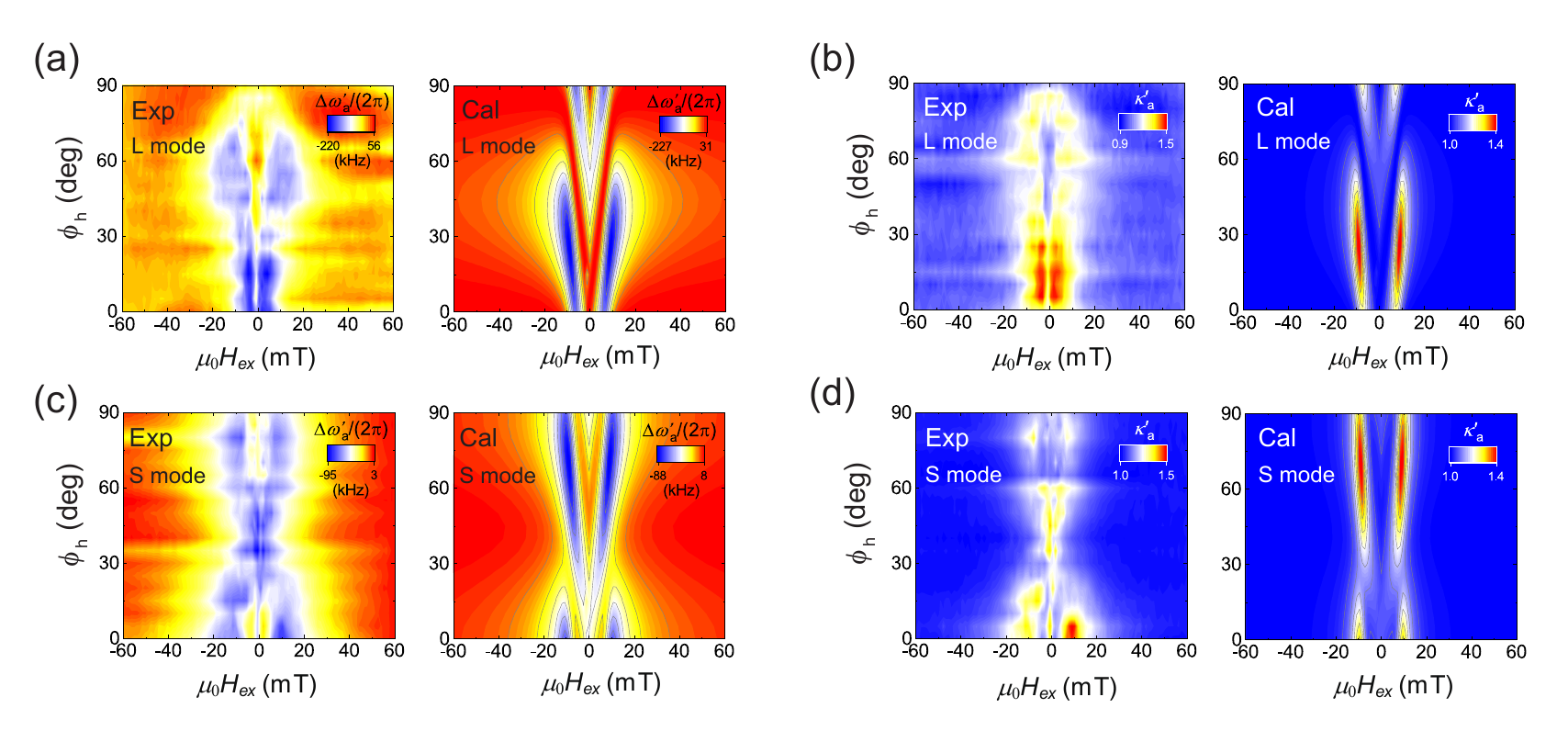}
		\vspace{0.0cm}
		\caption{Magnetoelastic modulation of acoustic cavity to the field orientation. \textbf{(a)-(d)} Field angle ($\phi_{\rm h}$) response of the resonant frequency shift ($\Delta \omega^\prime_{\rm a}$) and normalized acoustic damping ($\kappa^\prime_{\rm a}$) of L and S modes, where the normalized acoustic damping is defined as $\kappa_{\rm a}^{\prime} = \kappa_{\rm a}/\kappa_{\rm a0}$ where $\kappa_{\rm a0}$ is the value at $\left|\mu_0 H_{\rm ex} \right|$ = 60 mT. The left and right panels of each figure are experimental and calculated results. Here, we used $\kappa_{\rm a}^{\prime}$ but not $Q_{\rm a}^{\prime}$ in order to compare only the magnetoelastic back-action effects between different $\phi_{\rm h}$.}
		\label{fig 4}
		\vspace{-0cm}
	\end{center}
\end{figure*}
\hspace*{0.4cm}Acoustic phonons are possibly means for controlling magnons on-chip because of their micro-/nanometer wavelengths (similar to those of magnons), low-loss property, and negligible cross-talk \cite{li2021advances, zhang2016cavity, kikkawa2016magnon, berk2019strongly, an2020coherent, potts2021dynamical}. The pioneering studies on magnomechanical technology used surface acoustic wave (SAW) devices \cite{weiler_adfmr, dreher_adfmr, kobayashi_src, sasaki2017nonreciprocal, santos_magnomech2020, xu2020nonreciprocal, kawada2021acoustic, hatanaka_sawmag}. They succeeded in generating various magnetoelastic phenomena like acoustic spin pumping \cite{kobayashi_src, matsuo2013mechanical} and nonreciprocal transport \cite{sasaki2017nonreciprocal, xu2020nonreciprocal}. However, their large cavity structures are unsuitable for system integration and have difficulty taking full advantage of phonons \cite{hatanaka_sawmag}.\\
\hspace*{0.4cm}A phononic crystal (PnC) is a promising platform that enables acoustic phonons to be guided and trapped in a tiny wavelength-scale acoustic cavity and waveguide \cite{maldovan_nature, benchabane_pnc1, mohammadi_pnc1, otsuka_pnc, adibi_pncslab, baboly_pncslab2, hatanaka_hyperPnC}. Individual cavities are efficiently and finely driven and thereby can be used to control magnetic elements embedded in a PnC circuit. Moreover, the cavity sustains various spatial distributions of vibrational strains among multiple acoustic resonances, so it can be used to adjust the magnetoelastic effect \cite{hatanaka_pncmag_fem}. We consider that PnCs will enable us to make full use of phonons in magnomechanical technology.\\
\hspace*{0.7cm}Here, we demonstrate a PnC-based magnomechanical system sustained by cavity-waveguide coupled systems, as shown in Fig. 1(a). The nanomechanical vibrations confined in the cavity, excited through the waveguide, generate spin waves (magnons) in a nickel (Ni) film placed on its surface via magnetostriction. The acoustic spin pumping reversibly induces a frequency shift and damping modulation of the cavity resonances. Moreover, the magnon-phonon interaction can be tailored by selectively driving an appropriate cavity mode with specific strain distributions. PnC cavity magnomechanics is useful for on-chip control of magnons and phonons as well as their hybridized states (magnon polaron), and shows promise for extending the capabilities of classical and quantum information technologies.\\
\hspace*{0.4cm}The PnC is fabricated in free-standing GaAs, as shown in Fig.1(b) (the details of the fabrication and structure are presented in the Methods.). It consists of a snowflake triangular lattice with full bandgaps between 0.5 and 0.8 GHz (Fig. 1(c) and 1(d)) \cite{hatanaka_hyperPnC, safavi_snowflake}. Acoustic waves at frequencies within the bandgap propagate in a line-defect waveguide and drive a line-defect cavity. Measuring the cavity's spectral response reveals three acoustic resonances (Fig. 1(e)). Their modal shapes exhibit complete confinement of the vibrations in the defect (top panels of Fig. 1(f)). Numerical calculations with the finite-element method (FEM) reproduce these modal shapes and verify the origin of the observed peaks. In this way, the PnC cavity can strongly confine vibrations that are remotely driven through the waveguide.\\
\hspace*{0.4cm}The Ni film on the cavity surface has a magnetization whose precession is acoustically excited via magnetostriction. The magnetostrictive force that induces the precession is divided into two components $\mu_{0}h^{\rm am}_{1}$ and $\mu_{0}h^{\rm am}_{2}$ on the $1$- and $2$-axis, whose definitions are given in Fig. 2(a). The equilibrium magnetization axis ($m_{3}$) can be decomposed into in-plane $x$ and $y$ components and is at an angle ($\phi_{\rm m}$) from the waveguide direction $x$. The out-of-plane magnetostrictive force ($\mu_{0}h^{\rm am}_{1}$) is negligibly small because out-of-plane shear strains such as $\epsilon_{\rm xz}$ and $\epsilon_{\rm yz}$ vanish in the Ni at that location, whereas in-plane force ($\mu_{0}h^{\rm am}_{2}$) can be expressed as \cite{dreher_adfmr}
\begin{equation}
	\mu_{0}h^{\rm am}_{2} = b_{\rm am} \{(\epsilon_{\rm xx}-\epsilon_{\rm yy})\sin{2\phi_{\rm m}} - 2\epsilon_{\rm xy} \cos{2\phi_{\rm m}})\},
\end{equation}
where $b_{\rm am}$ is the magnetostrictive coefficient and $\epsilon_{\rm ij}$ is the vibrational strain component. Thus, the magnetostriction is governed by two major factors, vibrational strain ($\epsilon_{\rm xx}$, $\epsilon_{\rm yy}$ and $\epsilon_{\rm xy}$) and the magnetization direction ($\phi_{\rm m}$), whose contributions to the system are theoretically investigated below.\\
\hspace*{0.4cm}The magnetization angle ($\phi_{\rm m}$) can be predicted from the magnetic free-energy density normalized by the saturation magnetization ($M_{s}$), given by \cite{dreher_adfmr}
\begin{equation}
	G=-\mu_{0} \bm{H_{\rm ex}} \cdot \bm{m} - B_{\rm u} (\bm{m} \cdot \bm{u})^{2} + B_{\rm d} m_{\rm z}^{2},
\end{equation}
where $\mu_{0} \bm{H_{\rm ex}}$ and $\bm{m}$ are the external magnetic field and unit vector of magnetization, respectively. The thin-film structure of the Ni results in a perpendicular magnetic anisotropy ($B_{\rm d}$). The angle $\phi_{\rm m}$ is determined by estimating the minimum of $G$. In this calculation, the in-plane anisotropic field ($B_{\rm u}$) and its unit vector ($\bm{u}$) are introduced so as to reproduce the experimental magnetization curve (Fig. 2(b)). For instance, the response of $\phi_{\rm m}$ as a function of $\mu_{0} H_{\rm ex}$ at $\phi_{\rm h}$ = 0$^{\circ}$ is shown in the top panel of Fig. 2(c). The magnetization is parallel to $\mu_{0}H_{\rm ex}$, i.e. $\phi_{\rm m} = 0^\circ$, when the field strength stays in the high field region $|\mu_{0} H_{\rm ex}|$ $>$ 20 mT. However, it undergoes a rotation to $\phi_{\rm m} = 90^{\circ}$ in the low field region $|\mu_{0} H_{\rm ex}| <$ 20 mT before reversing. In this way, the magnetization experiences a rotation and reversal while sweeping $\mu_{0} H_{\rm ex}$. This change in magnetization determines the magnon resonance frequency ($\omega_{\rm m} / (2\pi)$). The field response is shown in the bottom panel of Fig. 2(c). The frequency monotonically decreases with decreasing $\mu_{0} H_{\rm ex}$ and approaches the acoustic resonant frequency ($\omega_{\rm a} / (2\pi)$) at $|\mu_{0} H_{\rm ex}|$ = 7 mT, where the magnon-phonon frequency mismatch is minimized.\\
\hspace*{0.4cm}Another aspect determining the magnetostriction is the spatial distribution and direction of vibration strains. The observed acoustic resonances can be decomposed into three strain components (Fig. 2(d)-2(f) shows the spatial profiles of the longitudinal ($\left| \epsilon_{\rm xx} - \epsilon_{\rm yy} \right|$) and shear strain components ($\left| 2\epsilon_{\rm xy} \right|$) of the resonances). Shear (longitudinal) strain is dominant in the resonances at 0.581 GHz (0.593 GHz), whereas both strains are comparable at 0.552 GHz. The desired strain distributions can thus be generated by selectively actuating an appropriate resonance. Hereafter, the shear- and longitudinal-strain modes will be labeled S and L, while the mode with comparable strains will be labeled SL.\\
\hspace*{0.4cm}The magnetization dynamics and spatial strain profiles allow us to estimate the magnetostrictive coupling mode volume ($V_{\rm c}$), which characterizes the interaction efficiency. determined by $\phi_{\rm m}$ and the magnon-phonon spatial mode overlap (expression and derivation in the Methods). Figures 2(g)-2(i) show the simulated field dependence of $V_{\rm c}$ of the L, SL, and S modes at $\phi_{\rm h}$ = 0$^{\circ}$ and for various $\phi_{\rm h}$ between $0^\circ$ and $90^\circ$. For the L mode at $\phi_{\rm h} = 0^\circ$, $V_{\rm c}$ mostly vanishes in the high field region because $\phi_{\rm h} = \phi_{\rm m} = 0^\circ$ and $\left| 2\epsilon_{\rm xy} \right|$ is negligibly small. In contrast, it increases dramatically as $\mu_{0} H_{\rm ex}$ decreases below 20 mT. This enhancement at $\mu_0 H_{\rm ex} = 6$ mT is caused by the magnetization rotation from $\phi_{\rm m}$ = 0$^\circ$ to 45$^{\circ}$; thereby, the term $(\epsilon_{\rm xx} -\epsilon_{\rm yy}) \sin{2 \phi_{\rm m}}$ becomes non-zero. $V_{\rm c}$ temporarily returns to almost zero as $\phi_{\rm m}$ reaches 90$^{\circ}$ just before the magnetization reverses and then approaches the original value after the increase at $\phi_{\rm m} = 225^\circ$ and $\mu_0 H_{\rm ex} = -6$ mT. In contrast, the S mode exhibits the opposite field dependency, in which a finite $V_{\rm c}$ in the high field region is reduced in the low field region because the dominant magnetostrictive term is $2 \epsilon_{\rm xy} \cos{2 \phi_{\rm m}}$, not $(\epsilon_{\rm xx} -\epsilon_{\rm yy}) \sin{2 \phi_{\rm m}}$. Since the SL mode has almost equal contributions from both strain components, the variation with respect to $\mu_{0} H_{\rm ex}$ is moderate compared with the other two modes (Methods). As $\phi_{\rm h}$ increases, the susceptibility to $\mu_{0} H_{\rm ex}$ is cyclically modulated (right panels of Fig. 2(g)-2(i)). The change in direction of $V_{\rm c}$ while sweeping $\phi_{\rm h}$ is opposite between S and L modes, resulting from different dominant strain components. These results indicate that the cavity mode structures as well as the external field can be used to tune the magnetostrictive interaction.\\
\hspace*{0.4cm}To experimentally show the magnetostrictive interaction, the field response of the cavity resonance was measured at $\phi_{\rm h} = 0^\circ$. The observed resonant frequency shift ($\Delta \omega^\prime_{\rm a} / (2\pi)$) and quality factor ($Q_{\rm a}^{\prime}$) are plotted as a function of $\mu_{0} H_{\rm ex}$ in Fig. 3(a) and 3(b), where L and S modes are chosen for their opposite and distinct field susceptibilities. The response on the L mode exhibits dual dips at $|\mu_{0}  H_{\rm ex}|$ = 5 mT with a reduction in $\Delta \omega^\prime_{\rm a} / (2\pi)$ and $Q_{\rm a}^{\prime}$ in the low field region. This behavior can be understood from the theoretical formula,
\begin{equation}
	u_{\rm a}(\omega) = \frac{f_{\rm d}}{-\omega^{2} + \omega_{\rm a}^{2} -i \kappa_{\rm a} \omega + V_{\rm c} b^2_{\rm am} \chi_{\rm m}(\omega) /(\rho V_{\rm a})},
\end{equation}
where $\rho$, $V_{\rm a}$, $f_{\rm d}$ and $\omega$ are mass density, acoustic mode volume, driving force density and angular frequency, and $\chi_{\rm m}$ is magnetic susceptibility (Methods). The theoretical predictions are in agreement with the experimental results, indicating that the acoustic modulation is due to the magnon-phonon interaction, as described by our model. This interaction enables acoustic excitation of spin-wave oscillations in Ni, which exerts back-action force on the cavity resonance and tunes $\omega_{\rm a}^\prime$ and $Q_{\rm a}^\prime$.\\
\hspace*{0.4cm}Remarkably, ferromagnetic magnons were able to be driven by phonons in the tiny PnC cavity. The effective mode volume of the L mode is estimated to be $V_{\rm a} \approx 6.6$ $\mu$m$^3$ = 0.54 $\lambda^2 t$ with an acoustic wavelength $\lambda$ of 3.5 $\mu$m, which is $10^5 - 10^6$ times smaller than that of a SAW-based magnomechanical cavity system \cite{hatanaka_sawmag}. The tiny-energy vibrations are confined by the high $Q$ resonance, so they hardly affect surrounding systems, unlike conventional electromagnetic-wave-based magnetic devices. We believe that the PnC cavity-waveguide system would be a building block for a magnomechanical system and could be used as a local magnon driver and phonon modulator.\\
\hspace*{0.4cm}Strong back-action effects were also observed in the S mode, where shear strain dominate magnetostriction. A comparison with the effect in the L mode reveals the impact of the mode strain profiles on the interaction. Figure 3(b) shows that the dual dips in this mode at $\left| \mu_{0} H_{\rm ex} \right|$ = 10 mT are wider than those in the L mode because of the finite $V_{\rm c}$ in the high field region. In addition, a distinct center dip occurs around $\mu_0 H_{\rm ex} = -1$ mT, an effect of the magnetization rotation. The field response is distinctly different from that in the L mode and can be simulated with our model. Thus, our cavity geometry is also used to selectively drive the magnetization dynamics utilizing the difference in strain distributions.\\
\hspace*{0.7cm}To examine how the strain distribution affects the magnetoelastic modulation, the field responses of the acoustic resonant frequency ($\Delta \omega^\prime_{\rm a} / (2\pi)$) and damping rate ($\kappa_{\rm a}^\prime$) on the L and S modes were investigated at $\phi_{\rm m}$ ranging from $0^\circ$ to $90^\circ$ (Fig. 4(a)-4(d)). The magnetoelastic modulation regions in $\Delta \omega^\prime_{\rm a}$ and $\kappa_{\rm a}^{\prime}$ broaden as $\phi_{\rm h}$ increases from 0$^{\circ}$ to 60$^{\circ}$ in the L mode and shrink toward $\phi_{\rm h}$ = 90$^{\circ}$. In contrast, the S mode shows the opposite dependency, in which the modulation regions become narrow around $\phi_{\rm h}$ = 45$^{\circ}$. The theoretical calculations reproduce the experimental variations in both modes; here, the cyclic modulation while changing $\phi_{\rm h}$ is governed by $V_{\rm c}$, so the L and S modes show the opposite behaviors. Note that the experiment and theoretical model showed a moderate response to the magnetoelastic effect in the SL mode (Methods), supporting the validity of our model. We also found similar mode-sensitive magnetoelastic modulation in a PnC cavity with a different defect geometry (Methods). These results verify that the mode-tunable magnetostriction allows us to control magnomechanical states acoustically.\\
\hspace*{0.4cm}In conclusion, the ease of designing the PnC cavity and the small spatial leakage of its vibrations are useful for constructing integrated magnomechanical systems, in which microwave signal operations such as sensing, memory and processing are performed using magnons and phonons. Moreover, it is possible to build a tiny magnomechanical system of magnon wavelength size to efficiently control and read-out the magnetic state of a micro-/nano-ferromagnetic system such as a magnetic tunnel junction \cite{yuasa2004giant}. We believe that PnC cavity magnomechanics will expand the use of magnons and phonons and related technologies.\\
\section*{Acknowledgments}
This work was partially supported by JSPS KAKENHI(S) Grant Number JP21H05020.\\
\section*{Author contributions}
D.H. fabricated the device and performed the measurements and the data analysis. M.A. made theoretical model, and D.H. and M.A conducted the simulations with support from H. Y. and H.O.. D.H. and M.A. wrote the manuscript. All authors discussed the results during preparation of the paper.\\
%
%\bibliography{ref_spin_pnc}
%

%
\section*{Methods}
\begin{appendix}
\section{Fabrication and measurement}
\hspace*{0.4cm}The magnomechanical PnC was fabricated from GaAs (1.0 $\mu$m)/ Al$_{0.7}$Ga$_{0.3}$As (3.0 $\mu$m) heterostructure on a GaAs single-crystalline substrate. A periodic arrangement of snowflake-shaped air holes was formed by electron-beam lithography and dry etching. The GaAs layer, including the PnC lattice, was suspended by immersion in diluted hydrofluoric acid (5$\%$). The PnC geometry gives rise to a complete bandgap between 0.45-0.60 GHz and 0.65-0.71 GHz. The acoustic waveguide was constructed by removing one line from the lattice, thereby enabling single-mode propagation at frequencies within the bandgap \cite{hatanaka_hyperPnC}. The resonator (cavity) formed by removing two holes was located at one side of the waveguide and sustains multiple resonant vibrations. A ferromagnetic thin film of nickel (Ni) with a thickness of 50 nm was deposited on the surface of the cavity and holds spin-wave (magnon) resonances. 5-nm-thick gold (Au) film was deposited on the Ni layer for preventing oxidization. The free-standing PnC slab is sandwiched by inter-digit transducers (IDT) made from Cr (5 nm) / Au (35 nm). The IDT consists of 100 transducers arrayed with a period of 4.9 $\mu$m $\sim$ 5.2 $\mu$m.\\
\hspace*{0.7cm}Acoustic waves were piezoelectrically excited by applying alternating voltages to one IDT and optically measured with an optical interferometer (Neoark, MLD-101). The data on spectral response of the PnC cavity were obtained with a time-gating technique with a network analyzer (Keysight E5080A) to remove undesired electrical cross-talk signals. The acoustic resonance frequencies ($\omega^\prime_{\rm a}/ (2\pi)$) and quality factor ($Q^\prime_{\rm a}$) in Fig. 3(a) and 3(b) were obtained by making Lorentzian or exponential fittings to the spectral and temporal response results. The acoustic damping rates ($\kappa_{\rm a}$, $\kappa_{\rm a0}$) in Fig. 4(b) and 4(d) were estimated from both $\omega^\prime_{\rm a}$ and $Q^\prime_{\rm a}$. The displacement amplitudes of the resonant mode profiles depicted in Fig. 1(f) were collected through frequency down-conversion followed by filtering with a lock-in amplifier (Stanford Research Systems, SR844). All experiments in this work were performed in a moderate vacuum ($10 - 100$ Pa) and at room temperature.\\
\section{Theory of magnetoelastic dyanmics in a phononic crystal cavity}
The equation of motion of the acoustic mode is given by
\begin{equation}
	\begin{split}
		\rho\left(\ddot{u}_i+\kappa_a \dot{u}_i+\omega_{\rm a}^2u_i\right) &= \partial_{x_{k}} \sigma_{ik}\\
		&= \omega_{\rm a}^2 u_i+\sum_{k=\{x,y,z\}} M_{\rm s} \partial_{k} \partial_{\epsilon_{ik}} G_d
	\end{split}
\end{equation}
where
\begin{align}
	G_{\rm d}= \sum_{l,n=\{x,y,z\}} b_{\rm am} \epsilon_{ln}m_l m_n.
\end{align}
is the magnetoelastic energy density. The magnetoelastic force density in the second term on the right-hand side is given by
\begin{align}
	\begin{split}
		f_{{\rm am},i}\equiv \sum_{k}M_{\rm s} \partial_{k} \partial_{\epsilon_{ik}} G_{\rm d} &= M_{\rm s} b_{\rm am} \sum_{k,l,n} \delta_{li}\delta_{nk}\partial_{k} m_l m_n\\
		&= M_{\rm s} b_{\rm am} (\nabla\cdot m_i{\bm m}).
	\end{split}
\end{align}
In our setup, the magnetization is aligned in-plane of the Ni film and the field angle away from the $x$-axis is defined as $\phi_{\rm m}$. Therefore, the conversion between $xyz$-coordinate system and $123$-coordinate system is
\begin{align}
	\left(\begin{matrix}m_x\\m_y\\m_z\end{matrix}\right)=\left(\begin{matrix}0&-\sin\phi_m&\cos\phi_m\\0&\cos\phi_m &\sin\phi_m\\-1&0&0\end{matrix}\right)\left(\begin{matrix}m_1\\m_2\\m_3\end{matrix}\right)
\end{align}
\begin{align}
	m_{\rm x}=&-\sin\phi_{\rm m} m_2+\cos\phi_{\rm m} \\
	m_{\rm y}=&\cos\phi_{\rm m} m_2+\sin\phi_{\rm m}\\
	m_{\rm z}=&-m_1
\end{align}
with approximations $m_1, m_2 \ll m_3 \sim 1$. Accordingly, under shear strains $\epsilon_{iz}\approx \epsilon_{zi}\approx 0$, the magnetoelastic force density in equation (B3) becomes
\begin{equation}
	\begin{split}
		f_{\rm am,x} / (M_{\rm s} b_{\rm am}) =&\sum_k\partial_{k} m_{\rm x} m_k\\
		=&-\sin2\phi_{\rm m} \partial_x m_2 + \cos2\phi_{\rm m} \partial_y m_2+\mathcal{O}(m_i^2)
	\end{split}
\end{equation}
and
\begin{equation}
	\begin{split}
		f_{\rm am,y} / (M_{\rm s} b_{\rm am}) =&\sum_k\partial_{k} m_{\rm y} m_k\\
		=&\cos2\phi_{\rm m} \partial_x m_2+\sin2\phi_{\rm m} \partial_y m_2+\mathcal{O}(m_i^2)
	\end{split}
\end{equation}
where $f_{\rm am,x}$ and $f_{\rm am,y}$ are the $x$ and $y$ components of the magnetoelastic force. As a result, we can define a new magnetoelastic vector,
\begin{align}
	{\bm v}_{\rm am}\equiv \frac{{\bm f}_\mathrm{am}}{M_sb_\mathrm{am}}=\left(\begin{matrix}-\sin 2\phi_{\rm m}  &\cos2\phi_{\rm m} \\\cos2\phi_{\rm m}&\sin 2\phi_{\rm m}\end{matrix}\right) \nabla\Phi_{m_2}({\bm r})
\end{align}
where $\Phi_{m_2}({\bm r})$ is the normalized amplitude of magnons at position $r$. This magnetoelastic vector contributes to the coupling constant with the spatial integration of magnons. By redefining the acoustic mode as ${\bm u}({\bm r}, t)=U_a(t){\bm \Psi}({\bm r})$, we find that
\begin{equation}
	\begin{split}
		\ddot{U}_{\rm a}(t)+\kappa_a \dot{U}_{\rm a}(t)&+\omega_{\rm a}^2 U_{\rm a}(t)\\ & = M_{\rm s} b_{\rm am} \frac{\int\mathrm{d}{\bm r}{\bm \Psi}({\bm r})\cdot{\bm v}_\mathrm{am}}{\int\mathrm{d}{\bm r}\rho({\bm r})|{\bm \Psi}({\bm r})|^2} m_2(t)\\ & \equiv M_{\rm s} b_{\rm am}\frac{V_{\rm c0}}{m_\mathrm{eff}} m_2(t)
	\end{split}
\end{equation}
where
\begin{align}
	m_\mathrm{eff}\equiv \rho V_{\rm a} = \rho \int_V\mathrm{d}^3{\bm r} |{\bm\Psi}({\bm r})|^2
\end{align}
is the effective mass with $\max_{\bm r}\left[|{\bm \Psi}({\bm r})|^2\right]=1$ and
\begin{align}
	V_{c0}=\int_V\mathrm{d}^3{\bm r}{\bm \Psi}({\bm r})\cdot{\bm v}_{\rm am}.
\end{align}
is the magnetostrictive coupling mode volume. Note that $\Phi_{m_2}({\bm r})$ and $m_2(t)$ are non-dimensional variables.\\
\subsection{Equation of motion for magnon modes}
The previous work by Dreher $\it{et}$ $\it{al.}$ \cite{dreher_adfmr} derived the following equations of magnonic motion: 
\begin{align}
	\frac{\alpha}{\gamma} \dot{m}_1+(G_{11}-G_3) m_1-\frac{1}{\gamma} \dot{m}_2=&0
\end{align}
\begin{equation}
	\begin{split}
		&\frac{\alpha}{\gamma} \dot{m}_2 +(G_{22}-G_3) m_2+\frac{1}{\gamma} \dot{m}_1\\
		&=b_{\rm am} \left[\sin2\phi_{\rm m} (\partial_x u_{\rm x} -\partial_y u_{\rm y})-\cos2\phi_{\rm m}\left(\partial_y u_{\rm x}+\partial_x u_{\rm y}\right)\right].
	\end{split}
\end{equation}
To determine $\Psi_{m_i}$, we have to diagonalize equations (B14) and (B15),
\begin{align}
	\dot{m}_1=&G_1\alpha m_1+G_2m_2+\frac{\gamma}{1+\alpha^2}F_{\rm am}\\
	\dot{m}_2=&-G_1m_1+G_2\alpha m_2+\frac{\gamma\alpha}{1+\alpha^2}F_{\rm am}
\end{align}
where $G_i\equiv \gamma (G_3-G_{ii})/(1+\alpha^2)$ and $F_{\rm am} = b_{\rm am} \left[\sin2\phi_{\rm m} (\partial_x u_{\rm x} -\partial_y u_{\rm y})-\cos2\phi_{\rm m}\left(\partial_y u_{\rm x}+\partial_x u_{\rm y} \right)\right]$. Here $G_{11}=2B_{\rm d}$, $G_{22}=-2B_{\rm u} \sin^2(\phi_{\rm m}-\phi_{\rm u})$, and $G_3=-\mu_0 H_\mathrm{ex}\cos(\phi_{\rm m}-\phi_{\rm h})-2B_{\rm u} \cos^2(\phi_{\rm m}-\phi_{\rm u})$. Finally, we obtain
\begin{align}
	\left(\begin{matrix} \dot{m}_+\\ \dot{m}_- \end{matrix}\right)=\left(\begin{matrix}\lambda_+&0\\0&\lambda_- \end{matrix}\right)\left(\begin{matrix}m_+\\m_- \end{matrix}\right)+\frac{\gamma}{1+\alpha^2}F_{\rm am}P^{-1}\left(\begin{matrix}1\\\alpha \end{matrix}\right)
\end{align}
where
\begin{align}
	\lambda_\pm\equiv \frac{\alpha(G_1+G_2)\mp\sqrt{-2G_1G_2(2+\alpha^2)+(G_1^2+G_2^2)\alpha^2}}{2}
\end{align}
and
\begin{align}
	\frac{\gamma}{1+\alpha^2}F_{\rm am}P^{-1}(1,\alpha)^T\equiv (s_+,s_-)^T\frac{\gamma}{1+\alpha^2}F_{\rm am}.
\end{align}
$s_+$ and $s_-$ are defined as
%\footnotesize
\begin{widetext}
	\begin{align}
		s_+=&-i\frac{G_1(2+\alpha^2)-G_2\alpha^2+\alpha\sqrt{G_1^2\alpha^2+G_2^2\alpha^2 -2G_1G_2(2+\alpha^2)}}{2\sqrt{G_1^2\alpha^2+G_2^2\alpha^2-2G_1G_2(2+\alpha^2)}}\\
		s_-=&i\frac{G_1(2+\alpha^2)-G_2\alpha^2-\alpha\sqrt{G_1^2\alpha^2+G_2^2\alpha^2 -2G_1G_2(2+\alpha^2)}}{2\sqrt{G_1^2\alpha^2+G_2^2\alpha^2-2G_1G_2(2+\alpha^2)}}.
	\end{align}
	%\normalsize
	%\end{widetext}
	As a result, the equations of magnon motion becomes
	\begin{align}
		\dot{m}_\pm=\lambda_\pm m_\pm+ \frac{\gamma s_\pm}{1+\alpha^2}F_{\rm am}.
	\end{align}
	By decomposing the temporal and spatial parts as $m_\pm\to m_\pm(t)\Phi_\pm({\bm r})$, it can be expressed as
	%\footnotesize
	%\begin{widetext}
	\begin{align}
		\dot{m}_\pm(t) \Phi_\pm({\bm r})=\lambda_\pm m_\pm(t) \Phi_\pm({\bm r}) + \frac{b\gamma s_\pm}{1+\alpha^2}U(t)\left[\sin2\phi_{\rm m} \left(\frac{\partial \Psi_{\rm x}}{\partial x} - \frac{\partial \Psi_{\rm y}}{\partial y}\right)-\cos2\phi_{\rm m}\left(\frac{\partial \Psi_{\rm x}}{\partial y}+ \frac{\partial \Psi_{\rm y}}{\partial x}\right)\right]
	\end{align}
	%\normalsize
	%\end{widetext}
	and
	%\footnotesize
	\begin{align}
		\Phi_+({\bm r})=\Phi_-({\bm r})=\frac{1}{K}\left[\sin2\phi_{\rm m} (\partial_x \Psi_{\rm x} -\partial_y \Psi_{\rm y})-\cos2\phi_{\rm m} \left(\partial_y \Psi_{\rm x}+\partial_x \Psi_{\rm y} \right)\right]
	\end{align}
	%\normalsize
	%\end{widetext}
	where $K$ is a wavevector defined as $K\equiv \max_{\bm r}\left[\sin2\phi_{\rm m} (\partial_x \Psi_{\rm x} -\partial_y \Psi_{\rm y})-\cos2\phi_{\rm m} \left(\partial_y \Psi_{\rm x}+\partial_x \Psi_{\rm y}\right)\right]$. Accordingly, the diagonalized equations can be simplified to
	\begin{align}
		\dot{m}_\pm(t) =\lambda_\pm m_\pm(t) + \frac{b\gamma s_\pm K}{1+\alpha^2}U(t).
	\end{align}
	
	Importantly, we have the relationship 
	%\begin{widetext}
	\begin{equation}
		\begin{split}
			\left(\begin{matrix}m_+\\m_- \end{matrix}\right)=&\frac{1}{i}\left(\begin{matrix}
				-\frac{(G_1-G_2)\alpha-\sqrt{-2G_1G_2(2+\alpha^2)+(G_1^2+G_2^2)\alpha^2}}{2G_1}&-\frac{(G_1-G_2)\alpha+\sqrt{-2G_1G_2(2+\alpha^2)+(G_1^2+G_2^2)\alpha^2}}{2G_1}\\1&1 \end{matrix}\right)^{-1}\left(\begin{matrix}m_1\\m_2 \end{matrix}\right)
		\end{split}
	\end{equation}
	%\end{widetext}
	and thus, we use
	\begin{align}
		m_2\approx i( m_++m_-).
	\end{align}
	\subsection{Coupled mode equation}
	The above equations of motions of acoustic phonons and magnons lead to the following equation of motion of coupled modes,
	%\begin{widetext}
	\begin{align}
		\ddot{U}_{\rm a}(t)+\kappa_{\rm a} \dot{U}_{\rm a}(t)+\omega_{\rm a}^2 U_{\rm a}(t)	=&i M_{\rm s} b_{\rm am}\frac{V_{\rm c}}{Km_\mathrm{eff}}\left(m_+(t)+m_-(t)\right)\\
		\dot{m}_\pm(t) =&\lambda_\pm m_\pm(t) + \frac{b_{\rm am} \gamma s_\pm K}{1+\alpha^2}U_{\rm a}(t)
	\end{align}
	where the coupling mode volume has been redefined as $V_\mathrm{c}=\int_V\mathrm{d}^3{\bm r}{\bm \Psi}({\bm r})\cdot{\bm v}_{\rm am}$ and
	%\begin{widetext}
	\begin{align}
		{\bm v}_{\rm am}=\frac{1}{2}\left(\begin{matrix}
			-\nabla^2 \Psi_{\rm x}+\cos4\phi_{\rm m} \left(\partial_x^2\Psi_{\rm x}-\partial_y^2 \Psi_{\rm x}-2\partial_x \partial_y \Psi_{\rm y} \right)+\sin4\phi_{\rm m}\left(\partial_x^2\Psi_{\rm y}-\partial_y^2\Psi_{\rm y}+2\partial_x\partial_y \Psi_{\rm x}\right)\\
			-\nabla^2\Psi_{\rm y}+\cos4\phi_{\rm m}\left(\partial_y^2\Psi_{\rm y}-\partial_x^2\Psi_{\rm y}-2\partial_x\partial_y \Psi_x\right)+\sin4\phi_{\rm m}\left(\partial_x^2\Psi_x-\partial_y^2\Psi_x-2\partial_x\partial_y\Psi_y\right)
		\end{matrix}\right).
	\end{align}
	%\end{widetext}
	By solving equations (B29) and (B30) with an additional driving force, $f_{\rm d}$, we obtain the acoustic displacement amplitude ($U_{\rm a}(t) = u_{\rm a} e^{i \omega t}$) modulated by back-action from magnons, where
	%\begin{widetext}
	\begin{equation}
		u_{\rm a}(\omega) = \frac{f_{\rm d}}{-\omega^{2} + \omega_{\rm a}^{2} -i \kappa_{\rm a} \omega + V_{\rm c} b^2_{\rm am} \chi_{\rm m}(\omega)/(\rho V_{\rm a})},
	\end{equation}
	and the magnetic susceptibility is
	\begin{equation}
		\begin{split}
			\chi_{\rm m}(\omega) = -i \frac{M_{\rm s} \gamma}{1+\alpha^{2}}
			\left( \frac{s_{-}}{-i\omega -i\omega_{\rm m} +\frac{\kappa_{\rm m}}{2}} + \frac{s_{+}}{-i\omega +i\omega_{\rm m} +\frac{\kappa_{\rm m}}{2}} \right).
		\end{split}
	\end{equation}
	%\end{widetext}
	For the numerical calculation, we had to derive the value of $\sqrt{V_{\rm c}/m_\mathrm{eff}}$ at which the spatial function appears as the same order $|\Psi|^2$. This means that constant factors cancel out in their ratio. Magnetoelastic coupling coefficients ($b_{\rm am}$) of 5 T, 6T and 10 T were used for the simulations of the SL, S and L modes, respectively. The angle of the in-plane anisotropic field was set at $\Delta \phi_{\rm u} = \phi_{\rm u} - \phi_{\rm h}$ = 85$^\circ$, to reproduce the field response of the equilibrium magnetization as shown in Fig. 2(b). The table lists the other parameters of the calculations.\\
	\hspace*{0.4cm}By transforming into the rotating frame of the acoustic modes, i.e., $U_a(t)=Ae^{-i\omega_a t}$, the equation of motion of acoustic phonons becomes
	\begin{align}
		\dot{A}+\frac{\kappa_{\rm a}}{2}=-\frac{M_{\rm s} b_{\rm am}V_{\rm c}}{2\omega_{\rm a} K m_\mathrm{eff}}\left(m_+(t)+m_-(t)\right).
	\end{align}
	The coupled mode equation in the frequency domain is given by
	%\begin{widetext}
	\begin{align}
		&\left(\begin{matrix}-i\omega+\frac{\kappa_{\rm a}}{2}&+\frac{M_{\rm s} b_{\rm am}V_{\rm c}}{2\omega_{\rm a} Km_\mathrm{eff}}&\frac{M_{\rm s} b_{\rm am} V_{\rm c}}{2\omega_{\rm a} Km_\mathrm{eff}}\\
			-\frac{b_{\rm am}\gamma s_+ K}{1+\alpha^2}&-i\omega-\lambda_+&0\\
			-\frac{b_{\rm am} \gamma s_- K}{1+\alpha^2}&0&-i\omega-\lambda_-
		\end{matrix}
		\right)\left(\begin{matrix}U_{\rm a}(\omega)\\m_+(\omega)\\m_-(\omega)\end{matrix}\right)\nonumber
		\equiv& \left(\begin{matrix}-i\omega+\frac{\kappa_{\rm a}}{2}&G_{A+}&G_{A-}\\
			-G_{+A}&-i\omega-\lambda_+&0\\
			-G_{-A}&0&-i\omega-\lambda_-
		\end{matrix}
		\right)\left(\begin{matrix}U_{\rm a}(\omega)\\m_+(\omega)\\m_-(\omega)\end{matrix}\right)
		=\left(\begin{matrix}f_{\rm d}\\0\\0\end{matrix}\right).
	\end{align}
	%\end{widetext}
	Hence, the acoustic mode spectra is
	\begin{align}
		U_{\rm a}(\omega)=f_{\rm d}\left[-\frac{G_{A+}G_{+A}}{(\lambda_++i\omega)}-\frac{G_{A-}G_{-A}}{(\lambda_-+i\omega)}+(-i\omega+\frac{\kappa_{\rm a}}{2})\right]^{-1}.
	\end{align}
	Apparently, the symmetrized coupling strength is
	\begin{align}
		G_\pm\equiv \sqrt{G_{A\pm}G_{\pm A}}=b_{\rm am}\sqrt{\frac{M_{\rm s}\gamma  s_\pm}{2\omega_{\rm a}\left(1+\alpha^2\right)}}\sqrt{\frac{V_{\rm c}}{m_\mathrm{eff}}}.
	\end{align}
	%\end{widetext}
	%
	\begin{table*}[t]
		\begin{tabular}{lll}
			\hline\hline
			${\rho}$&mass density& 8900 kg/m$^{3}$\\
			$B_{\rm d}$&out-of-plane shape anisotropy& 0.2 T\\
			$B_{\rm u}$&in-plane magnetic anisotropy& 4 mT\\
			$\alpha$&Gilbert damping factor& 0.1\\
			$M_{\rm s}$&saturation magnetization& 370 kA/m\\
			$\gamma$&gyromagnetic ratio& 2.185$\mu_{\rm B}/\hbar$\\
			\hline\hline
		\end{tabular}
		\caption{Acoustic and magnetic parameters used in the calculations.}
	\end{table*}
\end{widetext}
\section{Magnetoelastic modulation of SL mode in a PnC cavity}
The left and right panels of Fig. 5(a) plot the resonant frequency shift ($\Delta \omega_{\rm a}^\prime /(2\pi)$) and quality factor ($Q_{\rm a}^\prime$) as functions of the bias field ($\mu_{0} H_{\rm ex}$) for the SL mode in the $\phi_{\rm h} = 0^\circ$. Their field responses reveal a dual dip structure due to the increased magnon-phonon interaction; the theoretical calculations (solid line) show the same dip. Compared with the other modes, the modulation magnitudes, i.e. dip depths, are small, because of the comparable contributions of the longitudinal ($\left| \epsilon_{\rm xx} - \epsilon_{\rm yy} \right|$) and shear strain components ($\left| 2\epsilon_{\rm xy} \right|$), and thus, the field dependency of the coupling ($V_{\rm c}$) is small. Figure 5(b) and 5(c) show the experimental (left) and simulated (right) field responses of $\Delta \omega_{\rm a}^\prime$ and $Q_{\rm a}^\prime$ while sweeping $\phi_{\rm h}$ from $0^\circ$ to $90^\circ$. The field regions of the magnetoelastic modulation exhibit moderate variation with $\phi_{\rm h}$ compared with the S and L modes. Clearly, the magnetoelastic dynamics are a consequence of the spatial strain distribution of this mode.\\
\begin{figure*}[h]
	\begin{center}
		\vspace{-0.0cm}\hspace{-0.0cm}
		\includegraphics[scale=1]{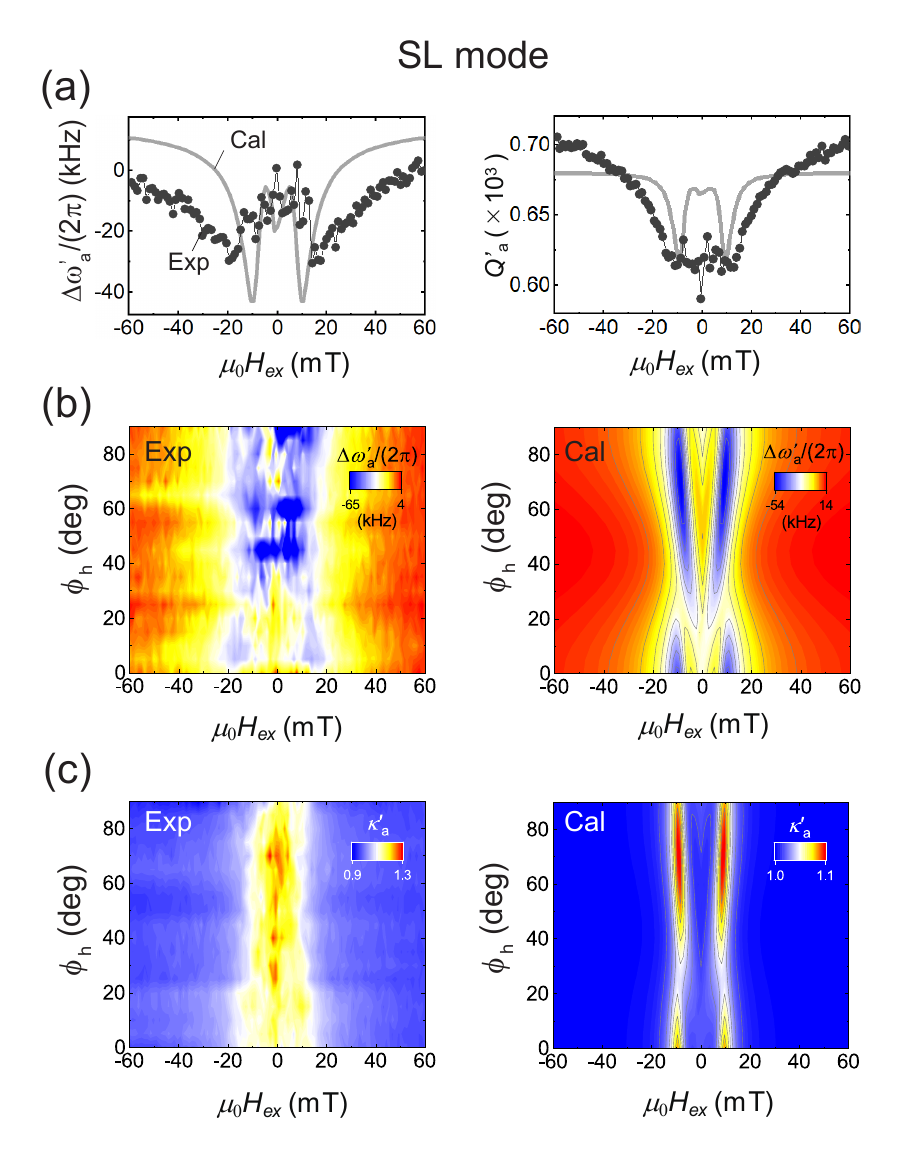}
		\vspace{0.0cm}
		\caption{Magnetoelastic modulation on an acoustic resonance of SL mode. \textbf{(a)} The field ($\mu_{0} H_{\rm ex}$) dependence of resonant frequency shift ($\Delta \omega_{\rm a}^\prime /(2\pi)$) and quality-factor ($Q_{\rm a}^\prime$) in $\phi_{\rm h}$ in the left and right panels respectively. \textbf{(b)} and \textbf{(c)} The $\phi_{\rm h}$ dependence of the frequency $\Delta \omega_{\rm a}^\prime /(2\pi)$ and damping $\kappa_{\rm a}^\prime$ versus $\mu_{0} H_{\rm ex}$ respectively. The experimental and calculated results are shown in the left and right panels.}
		\label{fig S_SL}
		\vspace{-0cm}
	\end{center}
\end{figure*}
\section{Magnetoelastic modulation on an PnC cavity with a three-holes defect}
A PnC cavity formed by removing three holes, shown in Fig. 6(a), holds two acoustic resonances at 0.579 GHz and 0.587 GHz with distinct modal shapes (Fig. 6(b) and 6(c)). These modes, labeled L and S, indicate that acoustic vibrations are confined in the defect (Fig. 6(d) and 6(e)). The numerically calculated spatial distributions of strains $\left| \epsilon_{\rm xx} - \epsilon_{\rm yy} \right|$ and $\left| 2\epsilon_{\rm xy} \right|$ are shown in the left and right panels of Fig. 7(a) and 7(b). Magnetoelastic coupling mode volume $V_{\rm c}$ as function of the bias field strength $\mu_{0} H_{\rm ex}$ at $\phi_{\rm h}$ = 0$^\circ$ is plotted in the left panels of Fig. 7(c) and 7(d) for the L and S modes. These field responses indicate that the two modes show opposite coupling dynamics. Similarly, the field angle evolution of the $V_{\rm c}$ response indicates the magnetoelastic modulation effect is sensitive to the field and magnetization orientation and the acoustic mode structure (right panels of Fig. 7(c) and 7(d)). Figure 8(a) and 8(c) indicate the frequency shifts of the L and S modes (left) and simulated responses (right). Figure 8(b) and 8(d) plot the normalized acoustic damping rates (left) and the simulated responses (right). The magnetoelastic coupling coefficient used in the calculations was $b_{\rm am} = 7$ T for both modes.\\
\begin{figure*}[h]
	\begin{center}
		\vspace{-0.0cm}\hspace{-0.0cm}
		\includegraphics[scale=1]{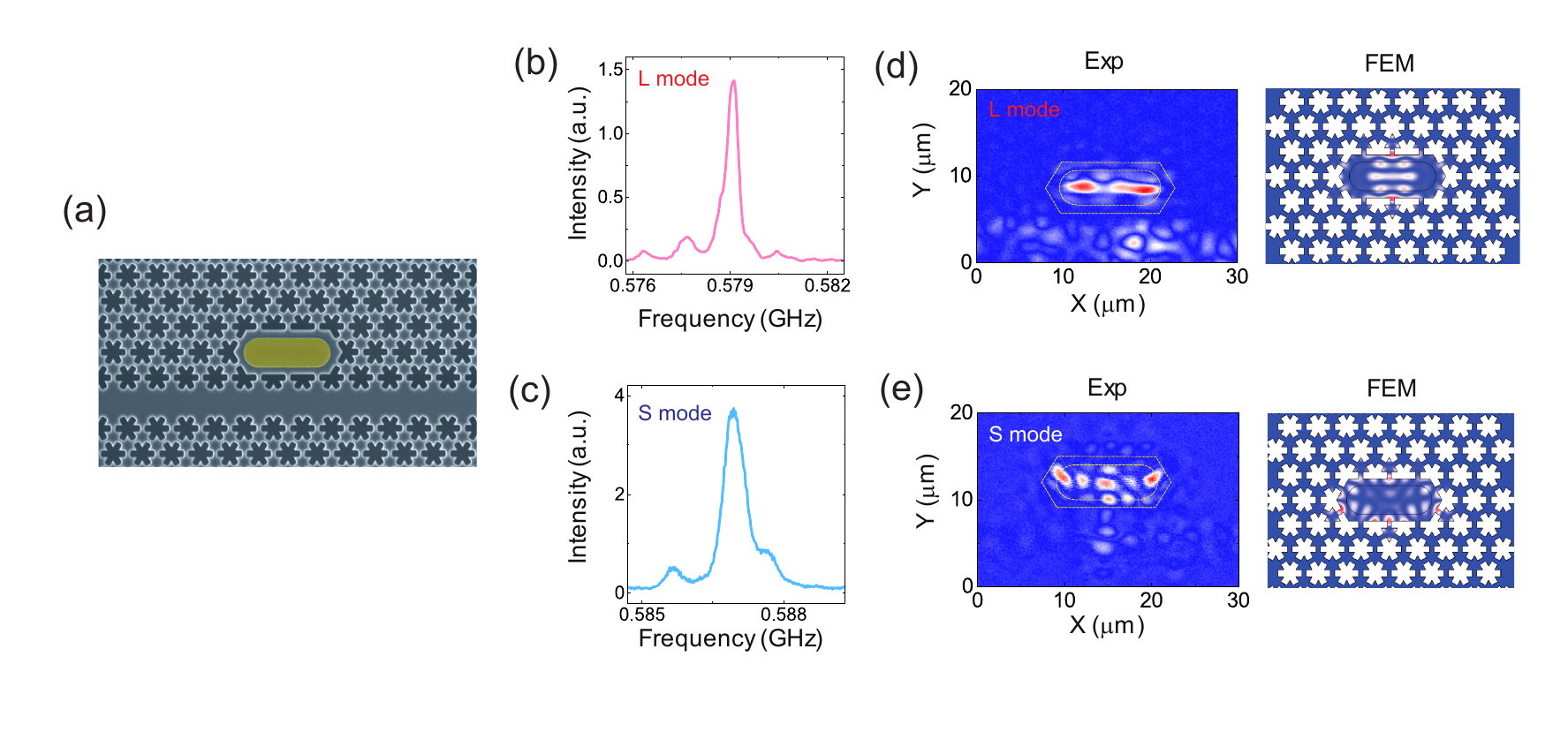}
		\vspace{0.0cm}
		\caption{Acoustic dynamics of PnC cavity with three hole defect. \textbf{(a)} SEM image of the PnC cavity coupled to a waveguide. A Ni film (yellow) is placed on the surface of the cavity. \textbf{(b)} and \textbf{(c)} Acoustic resonant spectrum of L and S modes. \textbf{(d)} and \textbf{(e)} Experimental (left) and calculated (right) spatial profiles of the resonant vibration amplitude in L and S modes. The defect region is highlighted by yellow dotted lines in the left panel.}
		\label{fig S_L3_1}
		\vspace{-0cm}
	\end{center}
\end{figure*}
\begin{figure*}[h]
	\begin{center}
		\vspace{-0.0cm}\hspace{-0.0cm}
		\includegraphics[scale=1]{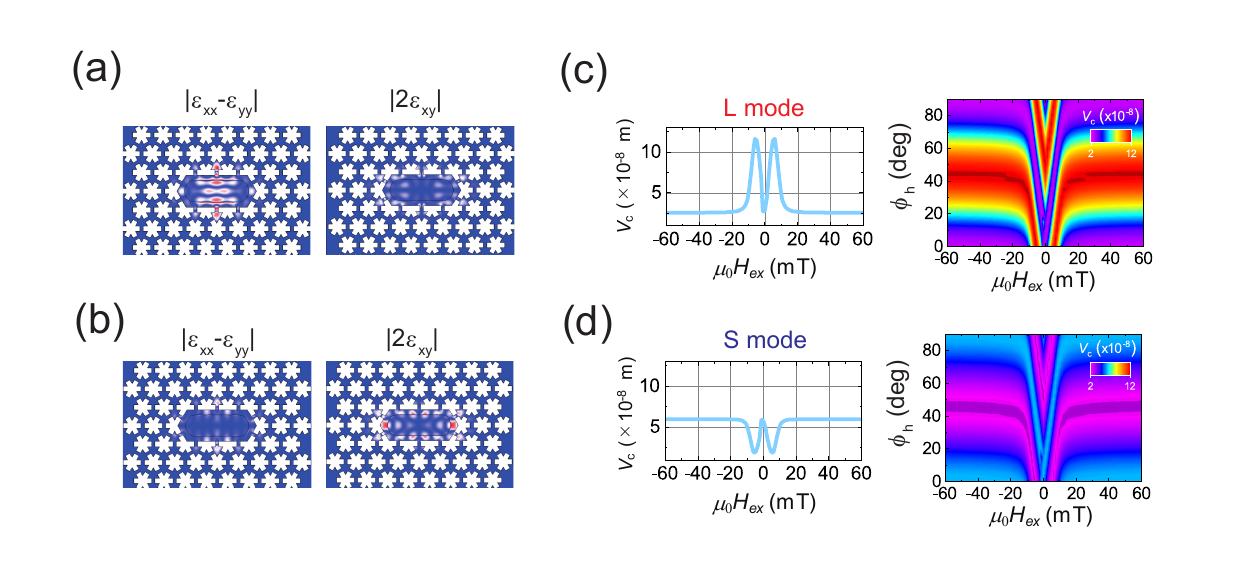}
		\vspace{0.0cm}
		\caption{Spatial strain profiles of cavity resonances and their magnetoelastic coupling mode volume. \textbf{(a)} and \textbf{b} Spatial distribution of longitudinal $\left| \epsilon_{\rm xx} - \epsilon_{\rm yy} \right|$ (left) and shear $\left| 2\epsilon_{\rm xy} \right|$ (right) strains on L and S modes, respectively. \textbf{(c)} and \textbf{(d)} Left: Simulated magnetoelastic coupling mode volume ($V_{\rm c}$) as a function of field strength ($\mu_{0} H_{\rm ex}$) at $\phi_{\rm h} = 0^\circ$. Right: Corresponding field angle ($\phi_{\rm h}$) dependence of $V_{\rm c} - \mu_{0} H_{\rm ex}$.}
		\label{fig S_L3_2}
		\vspace{-0cm}
	\end{center}
\end{figure*}
\begin{figure*}[h]
	\begin{center}
		\vspace{-0.0cm}\hspace{-0.0cm}
		\includegraphics[scale=1]{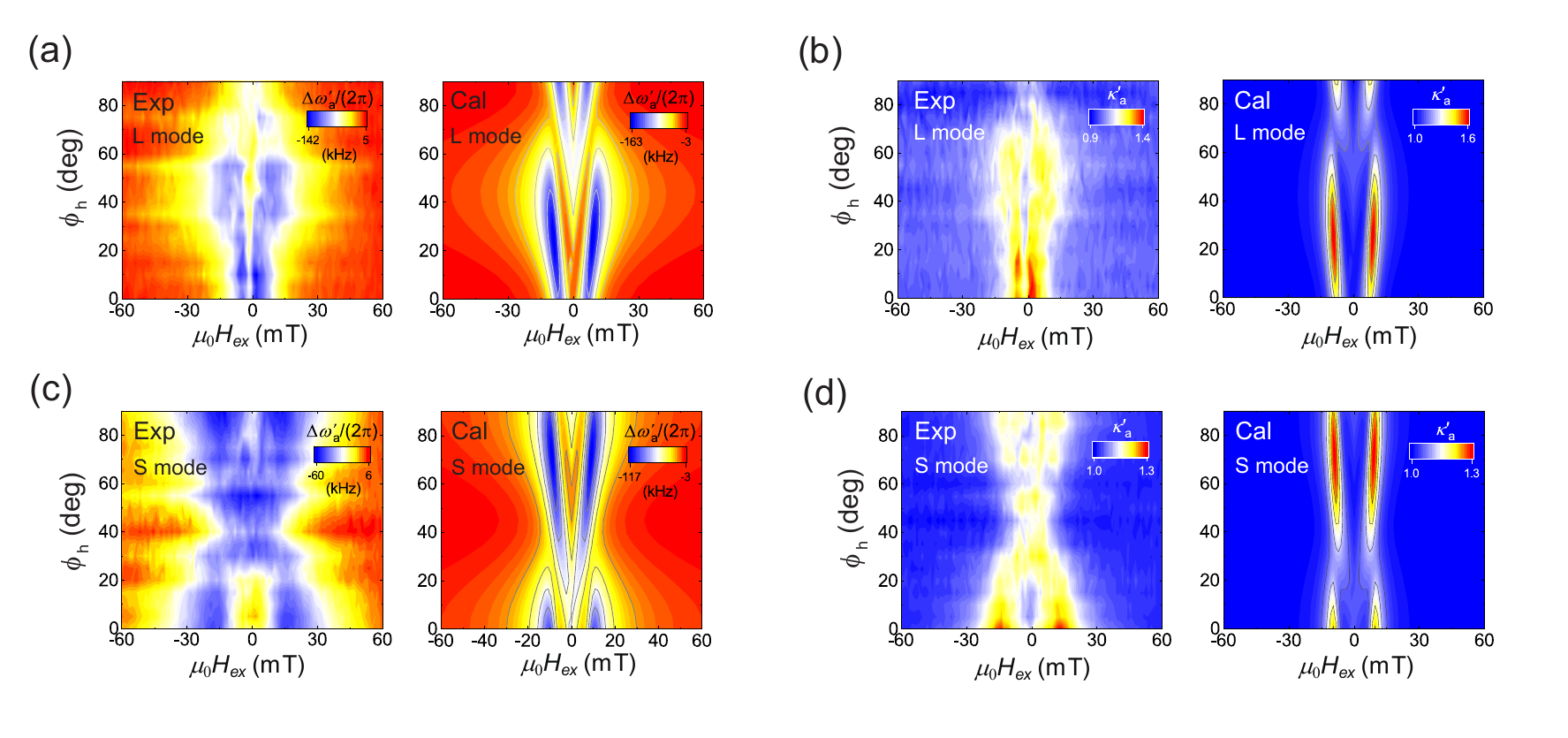}
		\vspace{0.0cm}
		\caption{Mode-sensitive magnetoelastic cavity dynamics. \textbf{(a)-(d)} Experimental (left) and simulated (right) $\phi_{\rm h}$ dependence of $\Delta \omega_{\rm a}^\prime /(2\pi)$ (a, c) and $\kappa_{\rm a}^\prime$ (b, d) as a function of $\mu_{0} H_{\rm ex}$ in L and S modes.}
		\label{fig S_L3_3}
		\vspace{-0cm}
	\end{center}
\end{figure*}
\end{appendix}
\end{document}